\documentclass[reprint,superscriptaddress,citeautoscript]{revtex4-1}

\usepackage[latin1]{inputenc}
\usepackage{graphicx}
\usepackage{fixmath}
\usepackage[colorlinks=true,linkcolor=black,%
citecolor=black,urlcolor=black]{hyperref}
\usepackage{amssymb}
\usepackage{amsmath}
\usepackage{amsthm}
\usepackage{bm}
\usepackage{psfrag}
\usepackage{mathrsfs}
\usepackage{dcolumn}
\usepackage[normalem]{ulem} %

\DeclareMathOperator{\erfc}{erfc}
\DeclareMathOperator{\erf}{erf}

\newcommand{\etal}{\emph{et al.}}

\newcommand{\oneline}[5]{
  #1 & #2 & #3 & #4 & #5  \\}
\newcommand{\soneline}[7]{
  #1 & #2 & #3 & #4 & #5 & #6 & #7 \\}

  \newcommand{\itap}{Institut f\"ur Theoretische und Angewandte Physik (ITAP),
  Universit\"at Stuttgart, Pfaffenwaldring 57, 70550 Stuttgart, Germany}
  \newcommand{\dprq}{D\'{e}partement de Physique and Regroupement
  Qu\'{e}b\'{e}cois sur les Mat\'{e}riaux de Pointe (RQMP),
  Universit\'{e} de Montr\'{e}al, C.P. 6128, Succursale Centre-Ville,
  Montr\'{e}al, Qu\'{e}bec, Canada H3C~3J7}

\begin{document}

\title{Influence of polarizability on metal oxide properties studied by molecular dynamics simulations}

\date{\today}

\author{Philipp Beck}
\affiliation{\itap}
\author{Peter Brommer}
\affiliation{\dprq}
\author{Johannes Roth}
\affiliation{\itap}
\author{Hans-Rainer Trebin}
\affiliation{\itap}

\begin{abstract}
 We have studied the dependence of metal oxide properties in
  molecular dynamics (MD) simulations on the polarizability of oxygen ions. 
  We present studies of both liquid and crystalline structures of silica (SiO$_2$), magnesia (MgO) and
  alumina (Al$_2$O$_3$). For each of the three oxides, two separately optimized sets of force fields were used: 
  (i) Long-range Coulomb interactions between oxide and metal ions
  combined with a short-range pair potential. (ii) Extension of force
  field (i) by adding polarizability to the oxygen ions. We show that while
  an effective potential of type (i) without polarizable oxygen ions
  can describe radial distributions and lattice constants reasonably
  well,  potentials of type (ii) are required to obtain correct values
  for bond angles and the equation of state. The importance of polarizability for
  metal oxide properties decreases with increasing temperature.
\end{abstract}

\maketitle

\section{Introduction}
\label{sec:intro}

Metal oxides are abundant in many technological applications.  Their
excellent insulating, thermal isolating and heat resisting properties
make them important components in microelectronics and semiconductor
engineering.  They also play a crucial role in nanotechnology and
nanoscience. Modelling these systems in classical atomistic
simulations requires effective potentials that describe the
interaction between the ions with reasonable accuracy. There exists a
wide selection of potential models with various degrees of
sophistication, from simple pair potentials to more intricate force
fields. 
All of these have to cope with the long-ranged nature of the Coulomb
interaction, which requires special treatment in large-scale
simulations. In this work, we will show that certain properties of
metal oxides cannot be adequately described in simulation with pure
pair potentials.

For crystalline silica  (SiO$_2$), Herzbach and Binder \cite{itapdb:Herzbach2005} 
already showed, that the polarizable
force field proposed by Tangney and Scandolo \cite{itapdb:Tangney2002}
is superior to both the 
BKS \cite{itapdb:Beest1990} and the fluctuating-charge DCG \cite{Demiralp1999} pair potential models.
Recently, polarization effects
on different properties of various molten fluorides, chlorides and
ionic oxides were described by Salanne and Madden \cite{Salanne2011}. The authors
illustrated the impact of polarizability in several exemplarily selected systems of these material classes and
predicted the general importance for structural and dynamic properties.

In this paper we present a systematic comparison of two types of
force fields for three different oxides, which only
differ by the presence of a polarizable term; the non-electrostatic
and Coulomb terms have in each case the same functional form
for both force fields. Hence, differences in
molecular dynamics (MD) simulation applications can be attributed to
polarizability.  To obtain a conclusion with a high degree of
universality, we present MD simulation studies of both liquid and
crystalline structures of silica (SiO$_2$), magnesia (MgO) and alumina
(Al$_2$O$_3$). Simulations were performed with the MD code
IMD \cite{IMD,IMD2}. 

For each of the three metal oxides, we used two separately optimized force fields:
\begin{enumerate}
 \item[(i)] Short range interactions of Morse-Stretch (MS) form combined with Coulomb 
    interactions between charged particles. 
 \item[(ii)] Extension of (i) by adding polarizability to the oxygen ions according to
    the Tangney-Scandolo (TS) potential model \cite{itapdb:Tangney2002}.
\end{enumerate}

The interactions between charges and/or induced dipoles are long ranged.
To handle these electrostatic forces correctly and efficiently, we applied the Wolf summation 
method \cite{itapdb:Wolf1999} in the potential generation as well as in simulations. In several
recent studies for metal oxides \cite{Brommer2010, Beck2011, Hocker2011}, linear scaling of
the computational effort in the number of particles could be achieved by using the Wolf summation without significant loss
of accuracy. 

Details of both TS model and Wolf summation are shown in Sec.~\ref{sec:force}, where also 
the two force field types are described in detail. In Sec.~\ref{sec:results},
whe present a systematic comparison of the two types of force fields for 
silica (SiO$_2$), magnesia (MgO) and alumina (Al$_2$O$_3$). Discussion and conclusion are given 
in Sec.~\ref{sec:discussion} and Sec.~\ref{sec:conclusion}, respectively.

\section{Force fields}
\label{sec:force}

\begin{table*}
   \centering
   \begin{tabular}{c c c c c c c c}
\hline
\hline
\soneline{$\phi^{0}_{\text{S}}$~~~}{$D_{\text{Si}-\text{Si}}$}{$D_{\text{Si}-\text{O}}$}{$D_{\text{O}-\text{O}}$}{$\gamma_{\text{Si}-\text{Si}}$}{$\gamma_{\text{Si}-\text{O}}$}{$\gamma_{\text{O}-\text{O}}$}
\soneline{}{~0.100 032~}{~0.100 250~}{~0.076 883~}{~11.009 449~}{~11.670 530~}{~7.505 632}
\soneline{}{$\rho_{\text{Si}-\text{Si}}$}{$\rho_{\text{Si}-\text{O}}$}{$\rho_{\text{O}-\text{O}}$}{$q_{\text{Si}}$}{$q_{\text{O}}$}{}
\soneline{}{2.375 990}{2.073 780}{3.683 866}{1.799 475}{-0.899 738}{}
\hline
\soneline{$\phi^{0}_{\text{M}}$~~}{$D_{\text{Mg}-\text{Mg}}$}{$D_{\text{Mg}-\text{O}}$}{$D_{\text{O}-\text{O}}$}{$\gamma_{\text{Mg}-\text{Mg}}$}{$\gamma_{\text{Mg}-\text{O}}$}{$\gamma_{\text{O}-\text{O}}$}
\soneline{}{0.038 258}{0.100 261}{0.065 940}{9.108 854}{10.405 694}{7.962 500}
\soneline{}{$\rho_{\text{Mg}-\text{Mg}}$}{$\rho_{\text{Mg}-\text{O}}$}{$\rho_{\text{O}-\text{O}}$}{$q_{\text{Mg}}$}{$q_{\text{O}}$}{}
\soneline{}{3.384 000}{2.417 339}{3.448 060}{1.100 730}{-1.100 730}{}
\hline
\soneline{$\phi^{0}_{\text{A}}$~~}{$D_{\text{Al}-\text{Al}}$}{$D_{\text{Al}-\text{O}}$}{$D_{\text{O}-\text{O}}$}{$\gamma_{\text{Al}-\text{Al}}$}{$\gamma_{\text{Al}-\text{O}}$}{$\gamma_{\text{O}-\text{O}}$}
\soneline{}{0.002 164}{1.000 003}{0.000 018}{10.855 181}{7.617 923}{16.719 817}
\soneline{}{$\rho_{\text{Al}-\text{Al}}$}{$\rho_{\text{Al}-\text{O}}$}{$\rho_{\text{O}-\text{O}}$}{$q_{\text{Al}}$}{$q_{\text{O}}$}{}
\soneline{}{5.517 666}{1.880 153}{6.609 171}{1.244 690}{-0.829 793}{}

\hline
\hline
  \end{tabular}
\caption{Parameters for the potentials $\phi^{0}_{\text{S}}$, $\phi^{0}_{\text{M} }$ and $\phi^{0}_{\text{A} }$, given in IMD units set
eV, \AA~and amu.}
\label{tab:para}
\end{table*}

\subsection{Generation}

All force fields were developed with the program \emph{potfit} \cite{potfit,potfit2},
which generates effective interaction potentials solely from \emph{ab initio} reference
structures. The potential parameters are optimized by matching the resulting forces, energies,
and stresses to corresponding first-principles values with the force matching 
method \cite{itapdb:Ercolessi1994}. In contrast to directly deriving charges and polarizabilities 
from first principles (as for example done in \cite{Heaton2006}) or fixing charges to their formal values (as for example done in \cite{Marro2009}), 
these values are also to optimize in \emph{potfit} 
in order to optimally reproduce forces, energies and stresses. It was already shown in \cite{itapdb:Tangney2002, Beck2011, Hocker2011},
that such optimal force-fits yield highly accurate interaction potentials. This also implies that charges and polarizabilities
obtained in this way are empirical quantities which need not correspond perfectly to a physical charge or polarizability. 

All reference data used in this study were obtained with the plane wave code VASP \cite{Kresse1993, Kresse1996}. 
For the optimization, a target function
\begin{equation}
Z = w_{\text{e}} Z_{\text{e}} + w_{\text{s}} Z_{\text{s}} + Z_ {\text{f}}
\end{equation}
is minimized. Here, $Z_{\text{e}}$, $Z_{\text{s}}$ and $Z_{\text{f}}$ are the contributions of the
quadratic deviations of energies (e), stresses (s) and forces (f), respectively.
$w_{\text{e}}$ and $w_{\text{s}}$ are certain weights to balance the different amount of
available data for each quantity. Root mean quare (RMS) errors, $\Delta F_l$
(with $l =$ e, s, f), are defined as proportional to the square root of the corresponding $Z_l$.
Their magnitudes are independent of weighting factors, number, and sizes of reference structures. 
For the minimization of the potential parameters, a combination of a stochastic simulated 
annealing algorithm \cite{itapdb:Corona1987} and a conjugate-gradient-like deterministic
algorithm \cite{itapdb:Powell1965} is used. 
Details of the whole optimization approach can be found in Ref.~\onlinecite{Beck2011} or \onlinecite{potfit}.

\subsection{Wolf summation}

The electrostatic energies of a condensed system are commonly computed with the Ewald method \cite{Ewald1921},
where the total Coulomb energy of a set of $N$ ions, $U_{\text{qq}}$, is decomposed into two terms $U^r_{\text{qq}}$ 
and $E^k_{\text{qq}}$ by inserting a unity of the form $1=\erfc\!\left(\kappa r\right)+\erf\!\left(\kappa r\right)$
with the error function
\begin{equation}
  \label{eq:Wolf2}
\erf\!\left(\kappa r\right):=\frac{2}{\sqrt{\pi}}\int\limits_0^{\kappa r}\!dt\,e^{-t^2}.
\end{equation}
$r$ is the spacial coordinate. The short-ranged erfc term 
is summed up directly, while the smooth erf term is
Fourier transformed and evaluated in reciprocal space. This restricts
the technique to periodic systems. However, the main disadvantage is
the scaling of the computational effort with the number of particles
in the simulation box, which increases as
$O(N^{3/2})$ \cite{itapdb:Fincham1994}, even for the optimal choice of
the splitting parameter $\kappa$.

Wolf \etal ~\cite{itapdb:Wolf1999} designed a method with linear scaling properties $O(N)$ for Coulomb
interactions. By taking into account the physical properties of the system,
the reciprocal-space term $U^k_{\text{qq}}$ is disregarded. 
In addition, a continuous and smooth cutoff of the remaining screened
Coulomb potential $\tilde{U}_{\text{qq}}(r_{ij})=q_iq_j\erfc(\kappa r_{ij}) r_{ij}^{-1}$
is adopted at $r_c$ by shifting the potential so that it goes to zero
smoothly in the first two derivatives at $r=r_c$. The Wolf method for charges 
and its extension \cite{Brommer2010} to dipolar interactions is applied to all force fields the present
article deals with. A detailed description of the Wolf summation with focus on its application
to dipole contributions can be found in Refs. \onlinecite{Brommer2010}, \onlinecite{Beck2011},
and \onlinecite{Hocker2011}, where it was successfully applied to silica, magnesia,
and alumina. 

\subsection{Type (i): MS + charges}

The potential $\phi^{0}_{\mu}$ (with $\mu =$ S, M, A for silica,
magnesia and alumina respectively) consists of a short range part of
MS form, and a Coulomb interaction between charged
particles. 
The MS interaction between two atoms $i$
and $j$ at positions $\bm{r}_{i}$ and $\bm{r}_{j}$ has the form
\begin{align}
\label{eq:MS}
  \phi_{\text{MS},ij} &=
  D_{ij}\left[\exp[\gamma_{ij}(1-\frac{r_{ij}}{\rho_{ij}})] -
    2\exp[\frac{\gamma_{ij}}{2}(1-\frac{r_{ij}}{\rho_{ij}})] \right], 
\end{align}
with $r_{ij}=\left|\bm{r}_{ij}\right|$,
$\bm{r}_{ij}=\bm{r}_{j}-\bm{r}_{i}$ and the model parameters $D_{ij}$,
$\gamma_{ij}$ and $\rho_{ij}$, which have to be optimized.  The
Coulomb interaction between two atoms is $U_{\text{qq},ij}=q_i q_j
r_{ij}^{-1}$. The charge $q_{\nu}$ of each atom type (with $\nu =$ O,
Si, Mg, Al for oxygen, silicon, magnesium and aluminum respectively)
is determined during the potential optimization under the constraint
of charge neutrality. The long-range
Coulomb interactions are treated with the Wolf summation method
both in force field generation and simulation to obtain $\phi_{\text{qq},ij}$. 
The total interaction is obtained by summing over all pairs of atoms:
\begin{equation}
\label{eq:Energy}
\phi^{0} = \phi^{0}_{\text{MS}} + \phi^{0}_{\text{qq}} = \sum_{\substack{i,j\\i>j}}(\phi_{\text{MS},ij}+\phi_{\text{qq},ij}). 
\end{equation}
With \emph{potfit}, we generated effective interaction potentials $\phi^{0}_{\mu}$ for silica, magnesia and alumina.
Table~\ref{tab:para} shows the corresponding parameters.

\subsection{Tangney-Scandolo dipole model}

In the TS model \cite{itapdb:Tangney2002}, the polarizability $\alpha$ of the oxygen atoms is taken into account. 
The dipole moments depend on the local electric field of the
surrounding charges and dipoles. Hence, a self-consistent iterative
solution has to be found. In the TS approach, a dipole moment
$\bm{p}_i^{n}$ at position $\bm{r}_i$ in iteration step $n$ consists
of an induced part due to an electric field $\bm{E}(\bm{r}_i)$ and
a short-range part $\bm{p}^{\text{SR}}_i$ due to the short-range interactions
between charges $q_i$ and $q_j$. Following Rowley
\etal ~\cite{itapdb:Rowley1998}, this contribution is given by
\begin{equation}
  \label{eq:TS1}
\bm{p}^{\text{SR}}_i = \alpha \sum\limits_{j \neq i} \frac{q_j
  \bm{r}_{ij}}{r_{ij}^3}f_{ij}(r_{ij}) 
\end{equation}
with
\begin{equation}
  \label{eq:TS2}
f_{ij}(r_{ij}) = c_{ij} \sum\limits_{k=0}^4 \frac{(b_{ij}r_{ij})^k}{k!}e^{-b_{ij}r_{ij}}.
\end{equation}
$f_{ij}(r_{ij})$ was introduced \emph{ad hoc} to account for multipole effects of 
nearest neighbors and is a function of very short range. 
$b_{ij}$ is the reciprocal of the length scale over which the short-range
interaction comes into play, $c_{ij}$ determines amplitude and sign of this
contribution to the induced moment.
$\alpha$, $b_{ij}$, and $c_{ij}$ have to be optimized additionally. 
Together with the induced part, one obtains
\begin{equation}
  \label{eq:TS3}
  \bm{p}_i^n=\alpha_i\bm{E}(\bm{r}_i;\{\bm{p}_j^{n-1}\}_{j=1,N},
  \{\bm{r}_j\}_{j=1,N}) +  \bm{p}_i^{\text{SR}}, 
\end{equation}
where $\bm{E}(\bm{r}_i)$ is the electric field at position $\bm{r}_i$, which is
determined by the dipole moments $\bm{p}_j$ in the previous iteration
step. Due to its excellent performance and scaling properties (see Fig.~\ref{fig:scaling} and Ref.~\onlinecite{Kermode2010}),
the TS model was used for more than 50 publications in the past ten years. 

\subsection{Type (ii): (i) + dipoles}

Beside charges, electric dipole moments on each oxygen ion are taken into account using the TS model.  
In addition to the interaction between charges ($\phi_{\text{qq}}$), the Wolf summation method is also 
applied \cite{Brommer2010} to the interaction charge-dipole ($\phi_{\text{qp}}$)
and dipole-dipole ($\phi_{\text{pp}}$). This yields the total interaction 
\begin{equation}
\label{eq:Energy2}
\phi = \phi_{\text{MS}} + \phi_{\text{qq}} + \phi_{\text{qp}} + \phi_{\text{pp}}.
\end{equation}
In this work, we use the polarizable force fields for silica \cite{Beck2011}, magnesia \cite{Beck2011} and alumina \cite{Hocker2011}
presented in earlier publications. These potentials were also generated with \emph{potfit}.

It must again be emphasized that while the potentials
$\phi^0_\mu$ and $\phi_\mu$ share the same functional form for
short-range and Coulomb interactions, their potential parameters were
optimized individually. Otherwise, a comparison would only lead
to the trivial result: A potential where some contributions are
omitted is no longer accurate. 

\subsection{RMS errors and scaling properties}

Although the potentials $\phi_\mu$ have three more parameters than the $\phi^0_\mu$ (13 compared to 10),  they do not describe the reference data significantly better than the $\phi^0_\mu$. 
This is illustrated by the RMS errors, that are shown in Table~\ref{tab:rms}. Indeed, there is a trend favoring the polarizable potentials: seven of nine RMS error are better in the 
case of $\phi_\mu$. And in the case of silica and magnesia, the RMS errors are - on average - smaller for $\phi_\mu$ than for $\phi^0_\mu$. For alumina, however, it is the other way round. 
Thus, the RMS errors are only first indicators of the quality of the generated force field, but they are not able to denote for
the practicability of a potential model.

\begin{table}
   \centering
\begin{tabular}{ p{1cm} p{2cm} p{2cm} p{2cm} }
\hline
\hline
$\mu$  & RMS error    & $\phi^0_\mu$ & $\phi_\mu$ \\
\hline
S      & $\Delta F_e$ &  0.216605    &  0.192220 \cite{Beck2011}  \\
       & $\Delta F_s$ &  0.040983    &  0.034099 \cite{Beck2011}  \\
       & $\Delta F_f$ &  1.866665    &  1.621107 \cite{Beck2011}  \\
\hline       
M      & $\Delta F_e$ &  0.075370    &  0.116994 \cite{Beck2011}  \\
       & $\Delta F_s$ &  0.029595    &  0.022774 \cite{Beck2011}  \\
       & $\Delta F_f$ &  0.671468    &  0.529535 \cite{Beck2011}  \\    
\hline       
A      & $\Delta F_e$ &  0.139316    &  0.049172 \cite{Hocker2011} \\  
       & $\Delta F_s$ &  0.055651    &  0.027273 \cite{Hocker2011} \\    
       & $\Delta F_f$ &  0.203966    &  0.350653 \cite{Hocker2011} \\       
\hline
\hline
\end{tabular}
\caption{RMS errors from the optimization of the force fields $\phi_\mu$ and $\phi^0_\mu$ (with $\mu =$ S, M, A for silica,
magnesia and alumina respectively).} 
\label{tab:rms}
\end{table}

\begin{figure}
  \centering
  \includegraphics{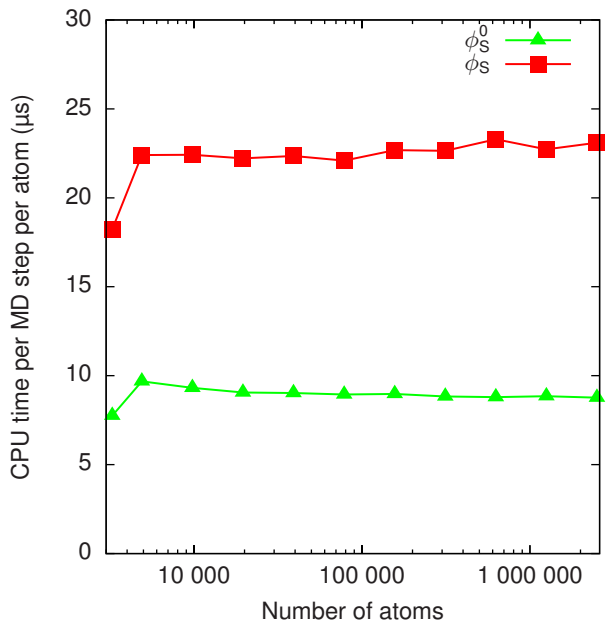}
 \caption{Scaling of computational effort with system size for $\phi^0_S$ and $\phi_S$. $\phi_S$ also scales linear with system size using
 Wolf summation, but is slower by a factor of about 2.6 compared to $\phi^0_S$. The scaling properties of simulations of magnesia and alumina
 are the same as those of silica: $\phi_{\mu}$ is slower than $\phi^0_{\mu}$ by a factor independent of system size.}
  \label{fig:scaling}
\end{figure}

All presented force fields are pure pair term potentials. Hence, simulations are less expensive compared to other approaches with
many-body potentials like a three-body interaction approach. Apart from that, considering polarizabiliy takes simulations time, because a new
self-consistent solution for all dipole moments has to be found in each MD time step. Taking dipoles into account slows down simulation
by a constant factor (in the case of silica about 2.6). The number of steps in the self-consistency loop is independent of 
the system size (Fig.~\ref{fig:scaling}). Both $\phi^0_\mu$ and $\phi_\mu$ yield linear scaling with the
number of particles due to the Wolf summation.  A comparison of the Wolf performance with two mesh-based methods can be found in Ref. \onlinecite{Beck2011}.
\section{Results}
\label{sec:results}

It is known from Ref.~\onlinecite{Beck2011} that force fields may also
yield qualitative results beyond the range for which they
were optimized, however such applications beyond the optimization
range should be closely verified. In the following, we focus the tests
on the range for the force
fields were trained, but we also show results outside this zone to
demonstrate the transferability of the potentials. 

\subsection{Microstructural properties}

\begin{figure*}
 \centering
\includegraphics[width=.325\linewidth]{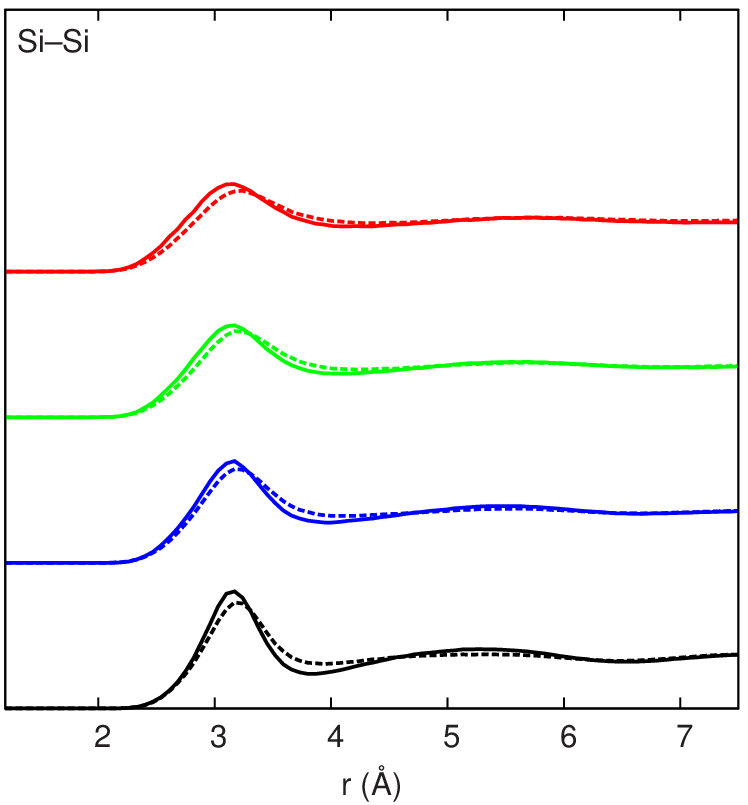}
\includegraphics[width=.325\linewidth]{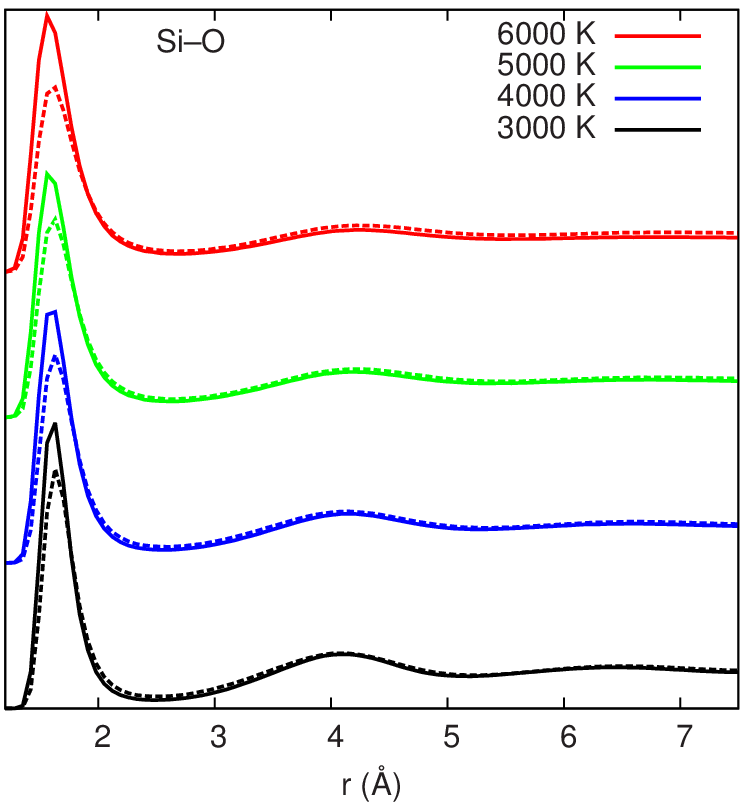}
\includegraphics[width=.325\linewidth]{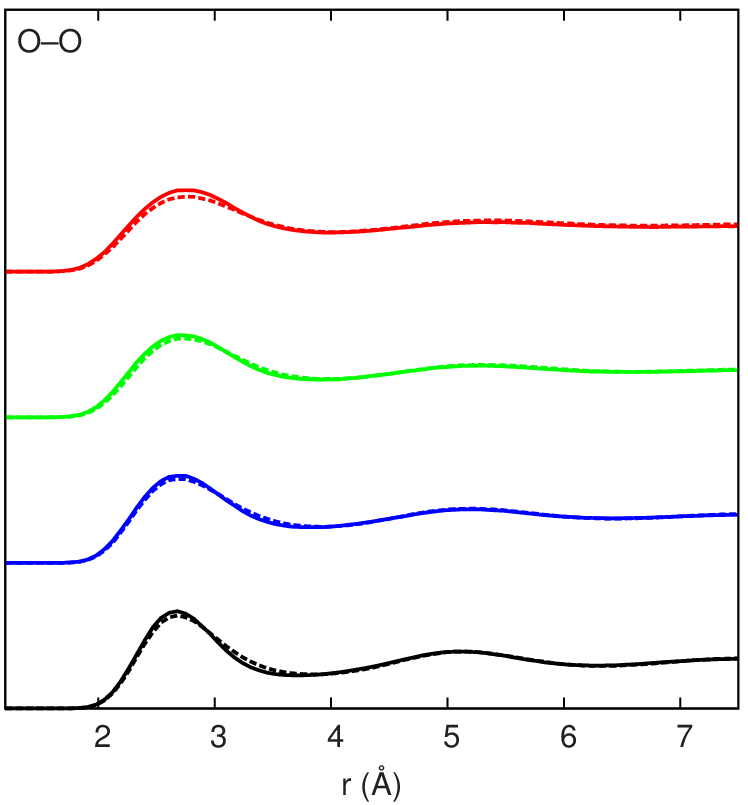}
\caption{Normalized radial distribution functions for Si--Si, Si--O and O--O in
  liquid silica at 3000--6000 K. The solid (dashed) curves belong to
   $\phi_{\text{S}}$ ($\phi^{0}_{\text{S}}$). }
 \label{fig:pair_silica}
\end{figure*}

\begin{figure*}
 \centering
\includegraphics[width=.325\linewidth]{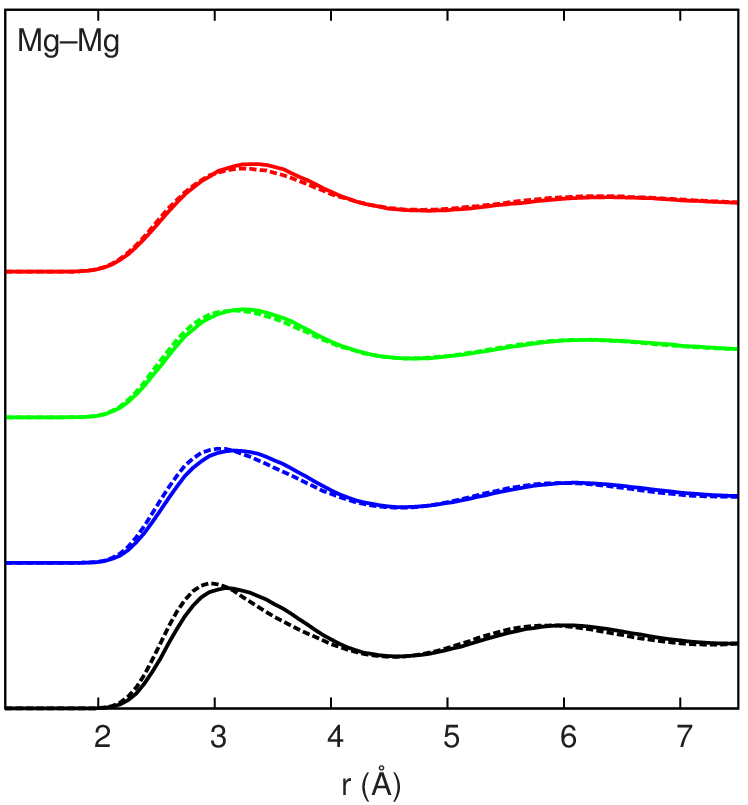}
\includegraphics[width=.325\linewidth]{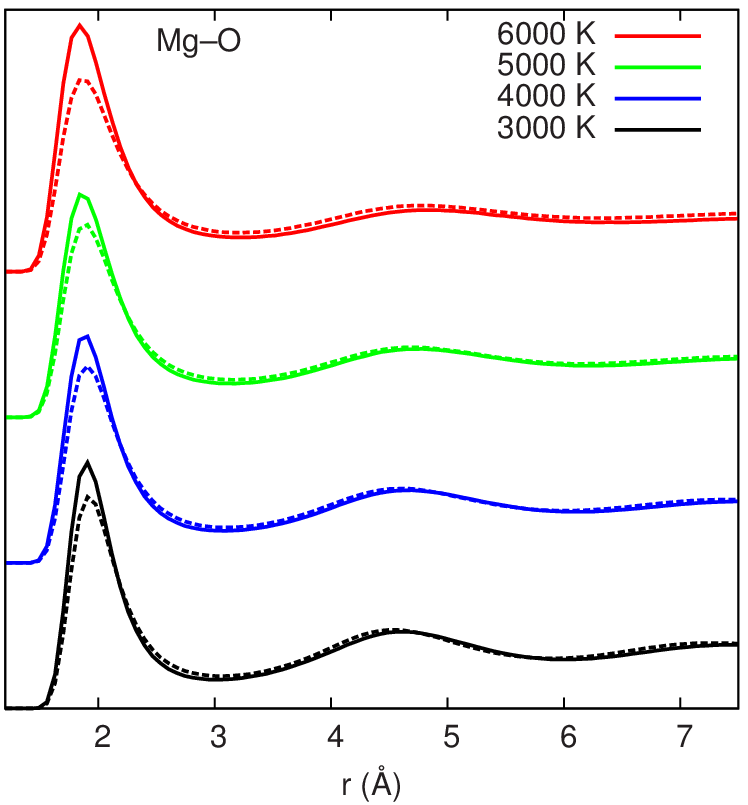}
\includegraphics[width=.325\linewidth]{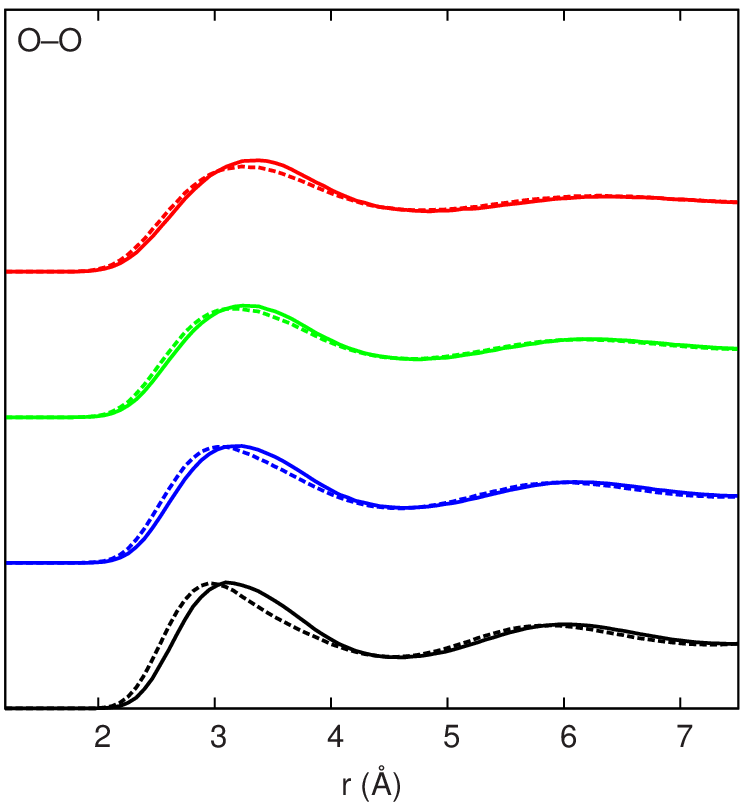}
\caption{Normalized radial distribution functions for Mg--Mg, Mg--O and O--O in
  liquid magnesia at 3000--6000 K. The solid (dashed) curves belong to
   $\phi_{\text{M}}$ ($\phi^{0}_{\text{M}}$). }
 \label{fig:pair_magnesia}
\end{figure*}

\begin{figure*}
 \centering
 \includegraphics[width=.65\linewidth]{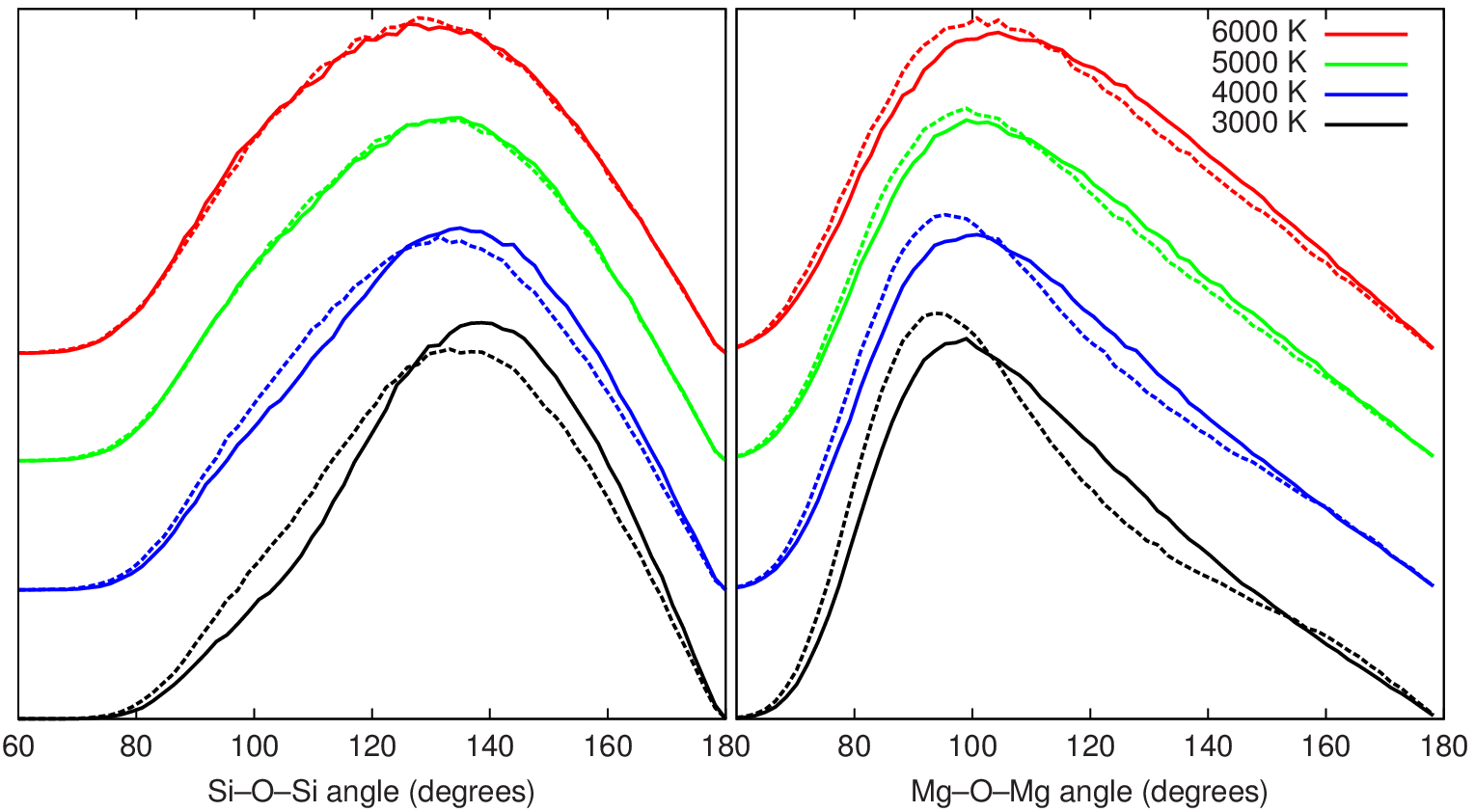}
 \includegraphics[width=.3285\linewidth]{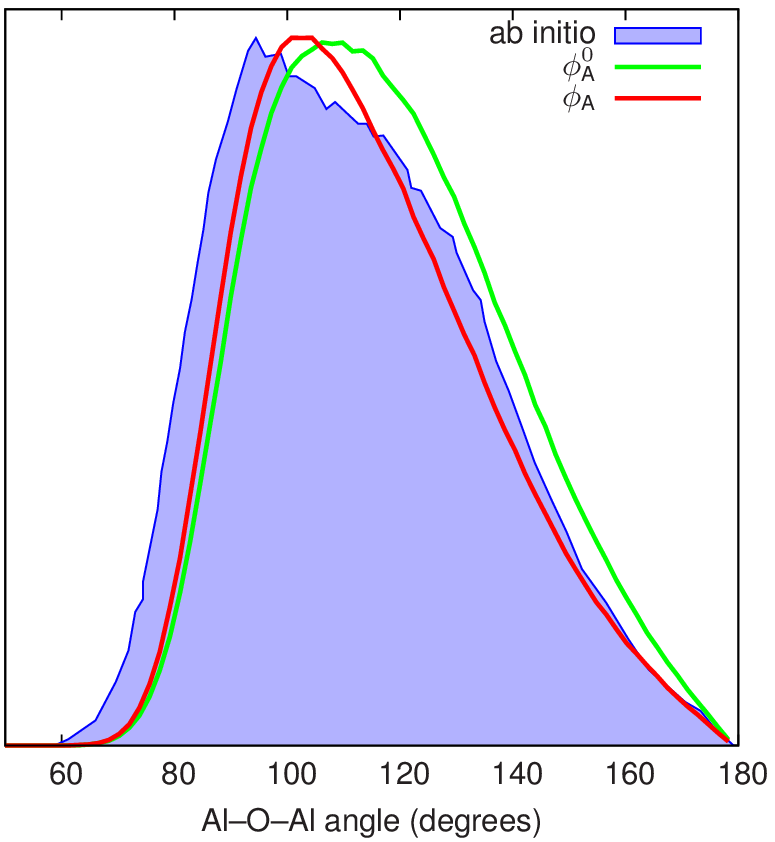}
\caption{Left: Normalized oxygen centered angle distributions in liquid
  silica and magnesia at 3000--6000 K for $\phi_{\text{S,M}}$ (solid
  curves) and $\phi^{0}_{\text{S,M}}$ (dashed
  curves). $\phi^{0}_{\text{S}}$ 
   yields a shoulder between around 80 and 130 degrees, whereas the band at around 135--170 degrees is underestimated. Similarly, $\phi^{0}_{\text{M}}$
   yields a shoulder between around 65 and 95 degrees, whereas the band at around 110--150 degrees is underestimated. Deviations decrease with increasing temperature.
   Right: The oxygen centered angle distribution in liquid alumina is shown at 3000 K for $\phi_{\text{A}}$ and $\phi^{0}_{\text{A}}$ and compared to an 
   \emph{ab initio} study \cite{Verma2011}. }
 \label{fig:angle}
\end{figure*}

\begin{table}
   \centering
\begin{tabular}{ p{2.5cm} p{1.5cm} p{1.5cm} l }
\hline
\hline
& ~a (\AA) & ~c (\AA) & E$_{\text{coh}}$ (eV) \\
\hline
 $\phi^{0}_{A}$ & 4.87 & 13.24  & ~34.71   \\
$\phi_{A}$ & 4.79  \cite{Hocker2011} & 12.97  \cite{Hocker2011}  & ~31.85  \cite{Hocker2011} \\
Ab initio  &  4.78  \cite{Hocker2011} & 13.05  \cite{Hocker2011}  & ~32.31  \cite{Hocker2011}   \\
Experiment & 4.75  \cite{Pear1991} & 12.99  \cite{Pear1991}  & ~31.8  \cite{Weast1983}   \\
\hline
\hline
\end{tabular}
\caption{Lattice constants and cohesive energy per Al$_2$O$_3$ unit of $\alpha$-alumina at zero Kelvin
obtained with $\phi_{A}$ and $\phi^{0}_{A}$ compared with \emph{ab initio}
results and literature data.} 
\label{tab:alumina}
\end{table}

\begin{table}
   \centering
  \begin{tabular}{l l l l l}
\hline
\hline
& \multicolumn{1}{l}{~$a$ (\AA)} & \multicolumn{1}{l}{~$c$ (\AA)} & \multicolumn{1}{l}{Si--O (\AA)} & \multicolumn{1}{l}{Si--O--Si} \\
 \hline
\oneline{$\phi^{0}_{S}$}{4.98}{5.47}{1.67}{139$^{\circ}$}
\oneline{$\phi_{S}$ \cite{Beck2011}}{5.15 }{5.50 }{1.65 }{148.5$^{\circ}$}
\oneline{Theory}{4.97 \cite{Gibbs2009}}{5.39 \cite{Gibbs2009}}{1.61 \cite{Gibbs2006}}{145$^{\circ}$ \cite{Gibbs2006}}
\hline
\hline
 \end{tabular}
\caption{Lattice constants, Si--O bond length and Si--O--Si angle of $\alpha$-quartz at 300 K compared with theoretical studies.} 
\label{tab:quartz}
\end{table}

First, the influence of polarizability on microstructural properties is illustrated. 
The radial distribution functions for liquid silica (4896 atoms) at 3000--6000 K
are, in each case, evaluated for 100 snapshots taken out of 100 ps MD
runs at the given temperature.
The averaged curves are given in Fig.~\ref{fig:pair_silica}. 
The curves obtained with $\phi^{0}_{\text{S}}$ are similar to the curves of $\phi_{\text{S}}$.
The existing slight deviations decrease with increasing temperature, so the polarizability
is more important for lower temperatures. For magnesia (see Fig.~\ref{fig:pair_magnesia}), the radial distribution functions show comparable behaviour: 
small deviations between  $\phi_M^0$ and $\phi_M$, that decrease with temperature. 
Apparently, the radial distribution in high temperature oxide melts
does not require polarizable oxide ions. 

A stronger influence of polarizability is observed on bond
angles. Fig.~\ref{fig:angle} (left) depicts the oxygen centered angle
distribution at 3000--6000 K in liquid silica and magnesia,
respectively. Both $\phi^{0}_{\text{S}}$ and $\phi^{0}_{\text{M}}$
overestimate the region of lower angles and underestimate the region
of higher angles. Potentials with electrostatic dipole moments yields the
correct shift of the distributions to slightly higher angles. Again,
deviations decrease with increasing temperature.  Although
$\phi_{\text{A}}$ and $\phi^{0}_{\text{A}}$ were optimized for
low-temperature crystalline structures, we probed the behavior for
liquid alumina at 3000 K. Fig.~\ref{fig:angle} (right) shows the
oxygen centered angle distribution in alumina compared to a recent
\emph{ab initio} study \cite{Verma2011}. Although the trend of
shifting angles to higher values by allowing for polarizability is not
reproduced in this case, polarizability yields a curve which is in
better agreement to \emph{ab initio} data. These results coincide with
Ref.~\onlinecite{Salanne2011}, where the authors stated that
polarization effects in ionic systems play an important role in
determining bond angles. 

The electrostatic dipole moments also influence crystalline structure parameters. The lattice constants of $\alpha$-alumina are given in Table~\ref{tab:alumina}.
$\phi_{\text{A}}$ yields an accurate agreement both with \emph{ab initio} and experimental data, whereas with $\phi^{0}_{\text{A}}$ the lattice
constants are overestimated (in each case around 2\%
deviation). However, both stabilize the trigonal crystal structure.
We also determined lattice constants, Si--O--Si angle and Si--O bond length for 
$\alpha$-quartz at 300 K (see Table~\ref{tab:quartz}), which is
outside the optimization range for the $\phi_{\text{S}}$ and
$\phi^{0}_{\text{S}}$ potentials. The average relative deviation of all parameters is for both potentials very similar 
(2.6\% for $\phi_{\text{S}}$ and 2.3\% for $\phi^{0}_{\text{S}}$).
On closer inspection, $\phi^{0}_{\text{S}}$ yields more accurate lattice constants, whereas $\phi_{\text{S}}$ better reproduces the
Si--O--Si angle and Si--O bond length. This is consistent with Ref.~\onlinecite{Salanne2011} and the results above concerning liquid metal oxides, 
where also the polarizability is more important for an improved description of bond angles rather than for atomic distances. 

\subsection{Thermodynamic properties}

\begin{figure}
  \centering
  \includegraphics{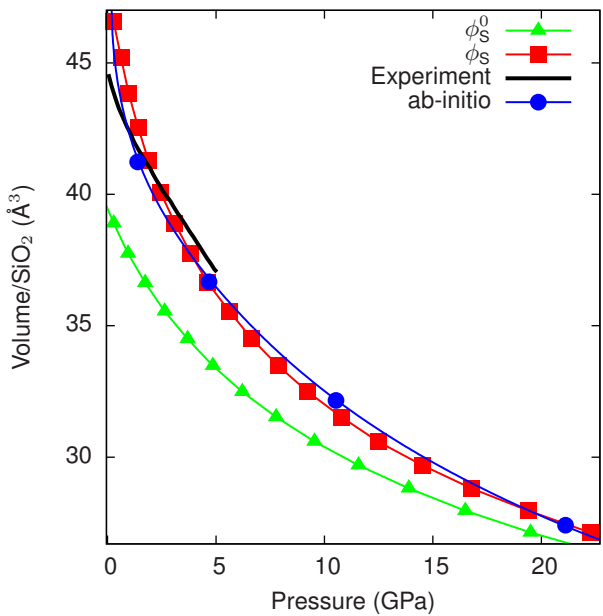}
 \caption{Equation of state of liquid silica at 3100 K for $\phi_{\text{S}}$ \cite{Beck2011} and $\phi^{0}_{\text{S}}$ compared with
   experiment \cite{Gaetani1998} and \emph{ab initio} calculations \cite{itapdb:Karki2007}. }
  \label{fig:eqstate_si}
\end{figure}

\begin{figure}
  \centering
  \includegraphics{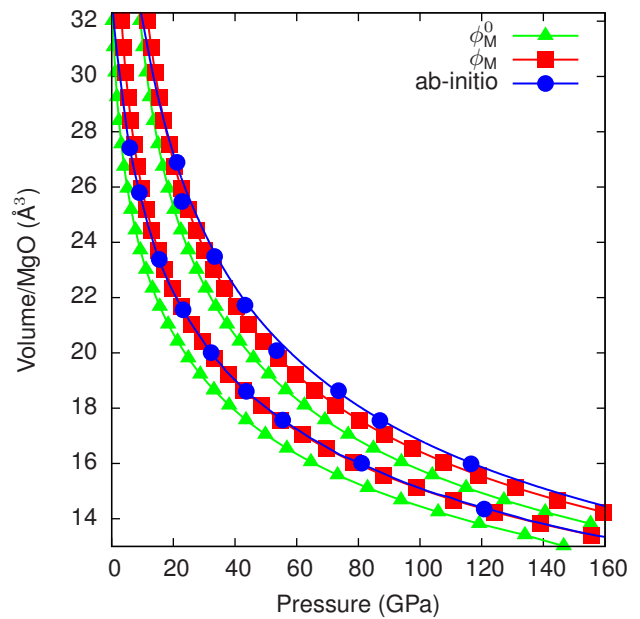}
 \caption{Equation of state of liquid magnesia at 5000 K (left three curves) and 10 000 K (right three curves) for $\phi_{\text{M}}$ and $\phi^{0}_{\text{M}}$ compared with
   \emph{ab initio} calculations \cite{Karki2006}. }
  \label{fig:eqstate_mg}
\end{figure}

For $\alpha$-alumina, we also calculated the cohesive energy (see Table~\ref{tab:alumina}). $\phi_{\text{A}}$ coincides with \emph{ab initio} and experimental
results, $\phi^{0}_{\text{A}}$ overestimates the cohesive energy (averaged deviation of 8.3\%). This clear deviation shows that electrostatic dipole moments
have to be taken into account, when probing macroscopical system properties.

To investigate the influence of polarizability on other thermodynamic properties,
we show the equation of state of liquid silica (3100 K, see Fig.~\ref{fig:eqstate_si}) 
and magnesia (5000 K and 10 000 K, see Fig.~\ref{fig:eqstate_mg}) respectively.
Pressures were obtained as averages along constant-volume MD runs 
of 10 ps following 10 ps of equilibration. The curves obtained with $\phi_{\text{S}}$ and 
$\phi_{\text{M}}$ coincide with \emph{ab initio} results as well as with experiment in the case of 
silica. The potentials $\phi^{0}_{\text{S}}$ and $\phi^{0}_{\text{M}}$, however, show a clear underestimation of
the volume, which illustrates the need for polarizability. 
The insufficiency of $\phi^{0}_{\text{M}}$ does not decrease with increasing temperature as in the 
case of microstructural properties. For the equation of state,
polarizability has to be taken into account
regardless of the simulation temperature.

\section{Discussion}
\label{sec:discussion}

Apparently, the equation of state of liquid oxides shows the most significant difference between $\phi^{0}_{\mu}$ and $\phi_{\mu}$; the potentials
without polarizability seem to lack a significant contribution to the pressure. This also applies to the (non-polarizable) BKS force field \cite{itapdb:Beest1990}, 
which underestimates pressure at fixed specific volume by a comparable amount (cf.~Ref.~\onlinecite{itapdb:Tangney2002}). 
The additional pressure in simulations with $\phi_{\mu}$ can however not directly be attributed to dipolar interactions. An analysis of the virial showed that $\phi_{\text{qp}}$ 
and $\phi_{\text{pp}}$ only contribute about 1.2\% of the total virial (and thus the pressure); the higher pressure for these systems results almost exclusively from
stronger MS and Coulomb contributions to the virial. As the atomic forces are described with comparable precision for both sets of potentials, this implies that the dipolar interaction 
is required to obtain correct pressures and forces simultaneously, especially as the polarizable force fields have higher absolute values of the atomic charges. 

When looking at the parameters of the force fields, it is noticable that in the $\phi^{0}_{\mu}$, the MS potentials are stronger at smaller atomic distances: The absolute value 
of the MS strengh $D_{ij}$ is higher and the stretch length $\rho_{ij}$ (cf.\ Eq.~\eqref{eq:MS}) is shorter in the non-polarizable potentials. This seems to indicate that in
this case, MS is required to describe the atomic interactions for nearest neighbours, while the dipolar interactions provide these contributions
for the force fields with polarization.

\begin{figure*}
 \centering
\includegraphics[width=.325\linewidth]{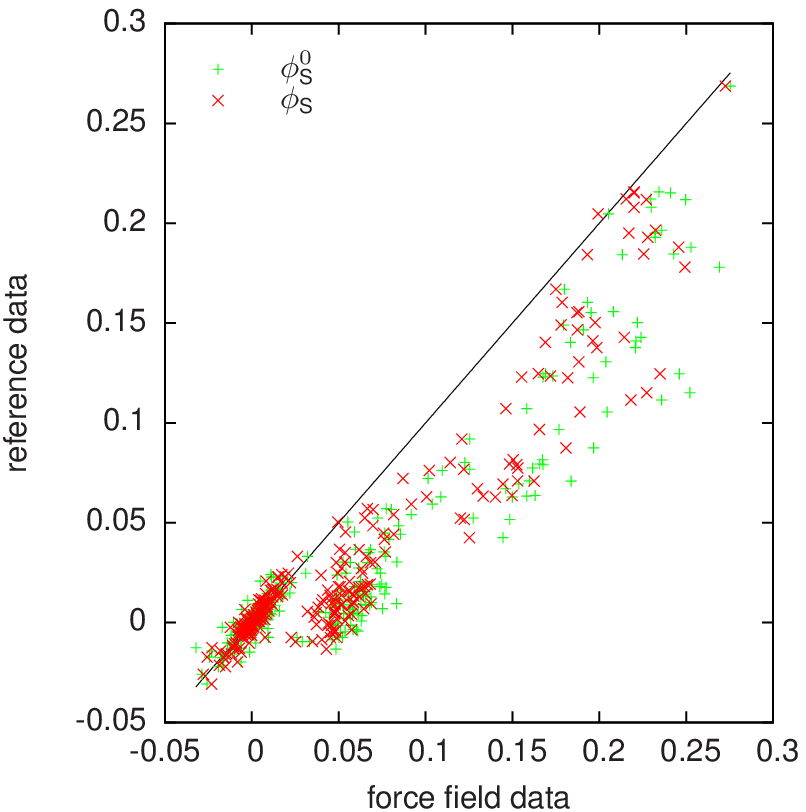}
\includegraphics[width=.325\linewidth]{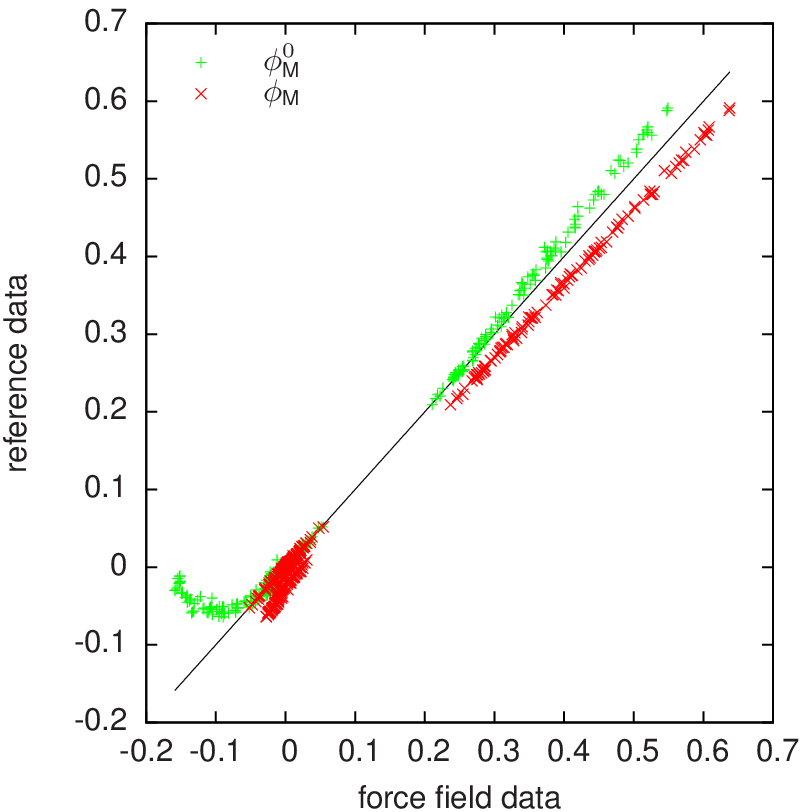}
\includegraphics[width=.325\linewidth]{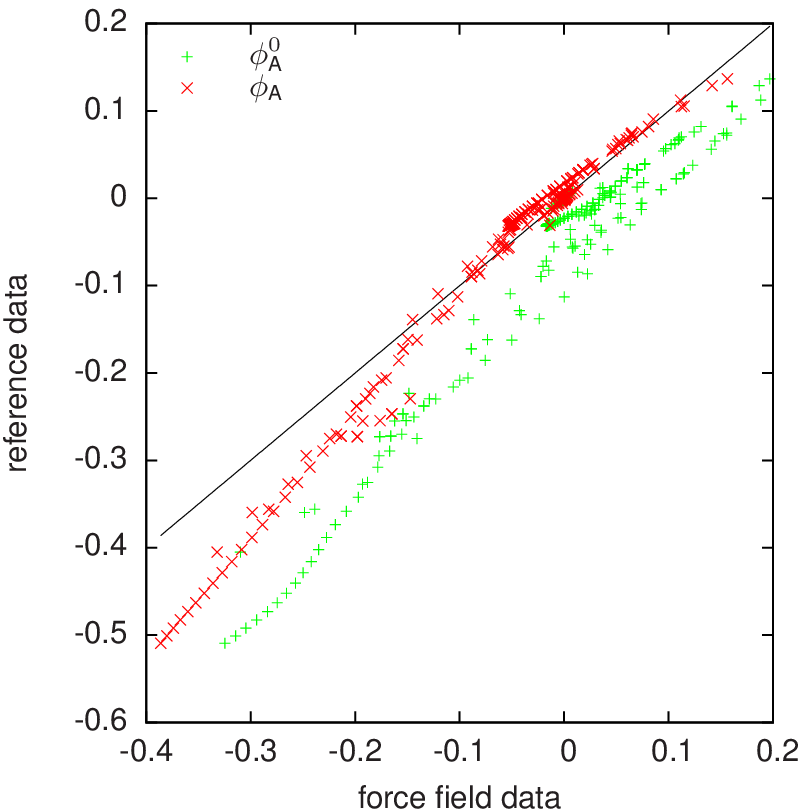}
\includegraphics[width=.325\linewidth]{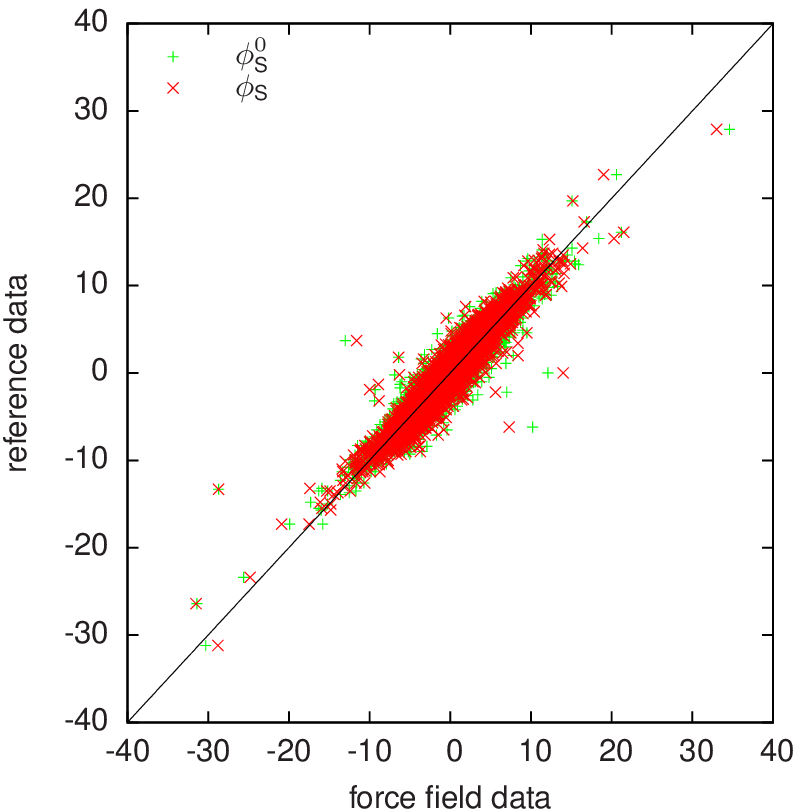}
\includegraphics[width=.325\linewidth]{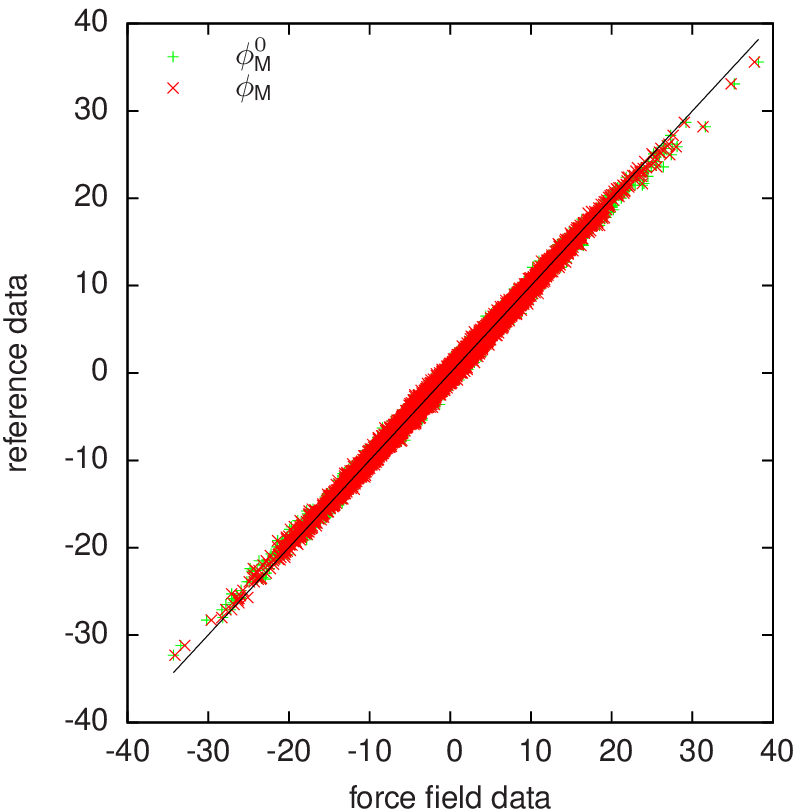}
\includegraphics[width=.325\linewidth]{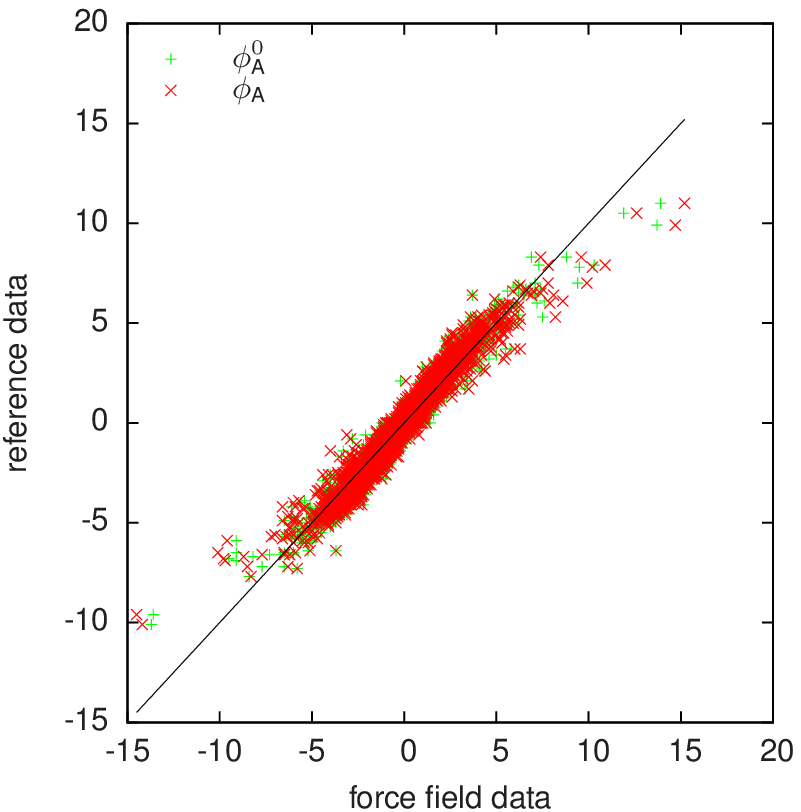}
\caption{Values computed by each potential (left silica, central magnesia, right alumina) plotted against its reference data value. A point placed on the bisecting line corresponds to
   perfect matching. Above: stress components of each configuration (in MPa); below: force components of each single atom (in meV/\AA).}
 \label{fig:scatter}
\end{figure*}

A better insight is uncovered by inspecting in detail, how accurate the reference \emph{ab initio} forces and stresses are reproduced by the particular force fields: Fig.~\ref{fig:scatter} 
depicts scatter plots, where for each single quantity (force component, stress component) the value computed by the potential is plotted against its reference data value. Hence, for perfect 
matching a point is placed on the bisecting line. As can be seen from the below graphs of Fig.~\ref{fig:scatter} showing the force components of each single atom, both force field types
$\phi_\mu$ and $\phi^0_\mu$ yield distributions scattered around the bisecting line. The only difference between $\phi_\mu$ and $\phi^0_\mu$ is how accurate the bisecting line is hit. 
In the top graphs, however, where the stress components of each configuration are depicted, clear deviations are uncovered: In the silica case, just a slightly worse matching of $\phi^0_S$
compared to $\phi_S$ is visible. But $\phi^0_M$ produces unnatural meanderings from the optimal matching at the left end of the graph (negative stress component values). The failure
becomes even more apparent in the alumina case, where $\phi^0_A$ underestimates the stresses over the whole data set. To sum up, the scatter plots predict less accuracy for the $\phi^0_\mu$,
when investigating system properties with a strong dependence on stress and pressure. 

\section{Conclusion}
\label{sec:conclusion}

In summary, we illustrated over a wide range the influence of polarizability on structural and thermodynamic properties in
liquid and crystalline systems of silica (SiO$_2$), magnesia (MgO) and
alumina (Al$_2$O$_3$), by comparing two distinct potentials for each material. 
As the functional form (but not the parameters) of the short-range part and the Coulomb term
are identical, the deviations between the results
of $\phi_{\mu}$ over $\phi^{0}_{\mu}$ could be associated with the additional dipole terms. 

We systematically investigated where the 
effects of electrostatic dipole moments are more important and how the impact arises from additional interaction mechanisms. 
The strongest influence of polarizability is observed in macroscopic thermodynamic properties as the equation of state and the cohesive
energy. Also on a microscopic scale, the role of dipoles is visible. However, the influences are more relevant for bond angle formation 
than for atomic distances in both liquid and crystalline structures. 
The influence always decreases with increasing temperature. 

Although the three presented metal oxides differ among each other in their stoichiometric configuration, the influence of polarizability is similar. Hence, it is expected that 
conclusions can be extended to other metal oxide systems. The results are not estimated to be limited to binary oxides as long as the same interaction mechanisms
are dominant. At present, the polarizable force field approach is applied to yttrium doped zirconia \cite{Irmler2012}, where a similar dependency of system properties on polarizability 
is observed. Statements concerning systems with different interaction mechanisms such as hydrogen bonds in water may exceed the present study. 

Previous comparisons of different oxide potentials \cite{itapdb:Herzbach2005,Salanne2011} have already shown that polarizable force fields are superior in many aspects. 
In the present work, we demonstrate directly, for which system properties dipole interactions are important and where the influence 
is negligible. Especially for the equation of state, simple pair potentials cannot reproduce the pressure-density relationship correctly. 

\acknowledgments
The authors thank Stephen Hocker for many helpful
discussions. Support from the DFG through Collaborative Research Centre 716, 
Project B.1 is gratefully acknowledged.

\end{document}